\documentclass[10pt,letterpaper]{article}
\usepackage{opex3}
\usepackage{amssymb,amsmath}

\usepackage{graphicx}

\begin{document}

\title{Atomic vapor quantum memory for a photonic polarization qubit}

\author{Young-Wook Cho$^*$ and Yoon-Ho Kim$^\dag$}

\address{Department of Physics, Pohang University of Science and Technology (POSTECH),\\Pohang, 790-784, Korea}

\email{$^*$choyoungwook81@gmail.com; $^\dag$yoonho72@gmail.com}

\homepage{http://qopt.postech.ac.kr}

\date{\today}

\begin{abstract}
We report an experimental realization of an atomic vapor quantum memory for the photonic polarization qubit. The performance of the quantum memory for the polarization qubit, realized with electromagnetically-induced transparency in two spatially separated ensembles of warm Rubidium atoms in a single vapor cell, has been characterized with quantum process tomography. The process fidelity better than 0.91 for up to 16 $\mu$s of storage time has been achieved.   
\end{abstract}

%\pacs{42.50.Gy, 32.80.Qk, 03.67.-a, 42.50.Ct}


%\maketitle

%%%%%%%%%%%%%%%%%%%%%%%%%%%%%%%%%%%%%%

%%%%%%%%%%%%%%%%%%%%%%%%%%%%%%%%%%%%%%%%%%%%
% Introduction 
%%%%%%%%%%%%%%%%%%%%%%%%%%%%%%%%%%%%%%%%%%%%

\section{Introduction}

Photons are one of the most convenient physical systems for encoding quantum information (or the qubit) and the photonic qubits (or the flying qubits) are known to be essential for implementing many novel ideas in quantum communication and  quantum computing \cite{kok,klm}. In recent years, a lot of research efforts have been focused on developing photonic qubit technologies essential for quantum information, including the photon source, encoding methods, propagation, storage, and detection. In particular, the photon storage device or the ability to store a photonic qubit has been known to be one of the most important problems, for example, in realizing linear optical quantum computing and long-distance quantum communication \cite{DLCZ, qm}.

Among the experimentally feasible approaches to quantum memory \cite{qm}, the method based on electromagnetically-induced transparency (EIT), which enables adiabatic transfer of a quantum state between an optical mode and a collective atomic excitation mode, stands out as the most promising approach to date \cite{eit,dark,Lukin}.   
Notably, storage and retrieval of a pulse of light have been demonstrated using the EIT effect in an atomic vapor cell \cite{Lukin}. Furthermore, the EIT-based storage/retrieval process is known to be able to preserve certain quantum properties of light. For instance, preservation of nonclassical properties of light during the storage/retrieval process in the EIT-medium has been reported for a conditional single-photon \cite{single1,single2}, filtered spontaneous parametric down-conversion photons \cite{SPDC}, and squeezed vacuum \cite{squeezed,honda}. In addition, it was recently demonstrated that thermal light can also be stored and retrieved in an EIT medium \cite{thermal}.

Since the photonic qubit is encoded with the superposition of a two-dimensional degree of freedom of a photon, e.g., polarization, path, time-bin, etc., a conventional $\Lambda$-type EIT scheme with a single dark state cannot store the photonic qubit which consists of two basis states. The use of two spatially separated atomic ensembles in a magneto-optical trap, however, allowed storage and retrieval of a heralded single-photon state superposed in two optical paths  \cite{mats,choi}. Ref.~\cite{choi} reports the storage/retrieval of the $45^\circ$ linear polarization state: the retrieved field, after 1 $\mu$s storage duration, shows the interference visibility of 0.91. Ref.~\cite{tanji} reports storage/retrieval of photon polarization states with two spatially overlapped cold atomic ensembles in an optical cavity: the polarization fidelity of 0.90 for 0.5 $\mu$s storage duration has been demonstrated. 
 
In this paper, we report an experimental demonstration of an atomic vapor quantum memory for the photonic polarization qubit. The atomic quantum memory for the polarization qubit is realized with the EIT effects of two spatially separated ensembles of warm Rubidium atoms in a single vapor cell. We have performed storage/retrieval experiments with arbitrary states of the photonic polarization qubit and quantified the performance of the atomic quantum memory with quantum process tomography. The atomic quantum memory for the polarization qubit exhibits the process fidelity better than 0.91 for up to 16 $\mu$s of storage time. Not only does our scheme exhibit high fidelity, long-storage of the polarization qubit, it does so in an inexpensive warm vapor cell rather than in cold atoms or optical cavities. This implies practicality of our scheme.

\section{Schematic of polarization qubit memory}

%%%%%%%%%%%%%%%%%%
\begin{figure}[tp]
\centering
\includegraphics[width=4.5in]{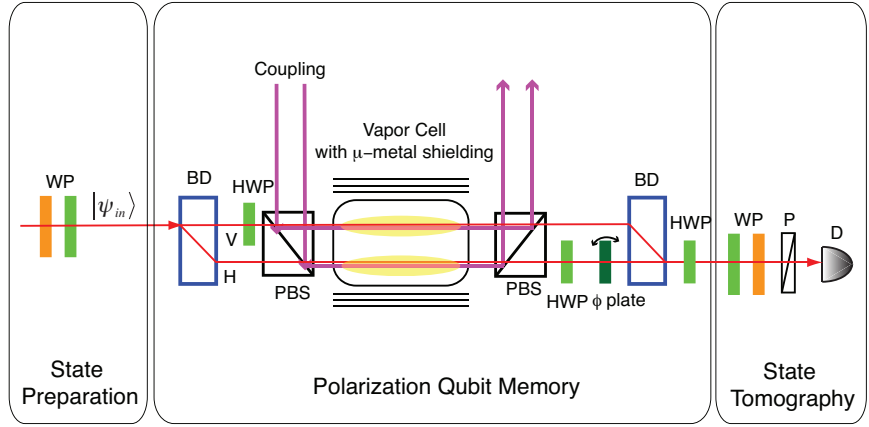}
\caption{Schematic diagram of the experiment. The photonic polarization mode is mapped onto two spatially separated atomic ensembles in a single warm vapor cell by the means of electromagnetically-induced transparency. The coupling fields are both vertically polarized. The polarization state of retrieved field is analyzed by a set of wave plates (WP) and a polarizer (P). BD: Beam displacer,  PBS: Polarizing beam splitter, HWP: Half-wave plate, $\phi$ plate: AR coated glass plate, D: single-photon detector. } \label{fig1}
\end{figure}
%%%%%%%%%%%%%%%%%%%%%% 

Let us first describe overall schematic. To store an arbitrary photonic polarization qubit, 
%%%%%%%%%%%%%%%%%%%%%%%%%%%%%%%
\begin{equation}
|\psi_{in}\rangle = \cos\theta |H\rangle + e^{i \phi} \sin\theta  |V\rangle, \label{state}
\end{equation}
%%%%%%%%%%%%%%%%%%%%%%%%%%%%%%%%
where $|H\rangle$ and $|V\rangle$ refer to horizontal and vertical polarization states, respectively, we map each of the polarization basis states to a warm atomic ensemble in a single atomic vapor cell via the EIT process. The two EIT media, each interacting with the $|H\rangle$ and $|V\rangle$ basis states for the photonic polarization qubit propagating in the $z$ direction, form the dark-state polaritons $\hat{\Psi}_{H}(z,t)$ and $\hat{\Psi}_{V}(z,t)$, respectively. For the two EIT media, we can write the two-state dark-state polariton as
%%%%%%%%%%%%%%%%%%%%%%%%%%%%%%%
\begin{equation}
\hat{\Psi}(z,t) = \cos\theta \hat{\Psi}_{H}(z,t) + e^{i \phi} \sin\theta \hat{\Psi}_{V}(z,t), \label{dsp}
\end{equation}
%%%%%%%%%%%%%%%%%%%%%%%%%%%%%%%%
where the dark-state polaritons $\hat{\Psi}_{H}(z,t)$ and $\hat{\Psi}_{V}(z,t)$ can be coherently manipulated by adiabatically changing the strengths of the coupling fields \cite{eit, dark}. The state of the atomic ensembles after the storage process can be considered as an atomic polarization state and we can map the state back to the photonic polarization qubit (or flying qubit) by turning on the coupling field.

The schematic of the experiment is shown in Fig.~\ref{fig1}. The probe field is initially prepared in an arbitrary polarization state, photonic polarization qubit $|\psi_{in}\rangle$, with a set of half-wave and quarter-wave plates (WP). A calcite beam displacer (BD) splits the probe beam into two spatial modes separated by 4 mm and BD is oriented such that the photons in the two modes are orthogonally polarized in $|H\rangle$ and $|V\rangle$. The vertical polarization probe mode, labeled as $V$ in Fig.~\ref{fig1}, is then rotated to horizontal polarization by using a half-wave plate (HWP). The two probe modes (both now horizontally polarized) are then combined with strong coupling fields (vertically polarized) at the polarizing beam splitter (PBS) and directed into the Rubidium vapor cell. The 75 mm long vapor cell is filled with isotopically pure \textsuperscript{85}Rb with 10 Torr Ne buffer gas and is wrapped with layers of $\mu$-metal to block the stray magnetic field. The temperature of the vapor cell was kept at 53$^{\circ}$C during the experiment and the corresponding optical depth $d$ was approximately $4$ \cite{note,phillips08}. 

After the vapor cell, the coupling beams are separated from the probe fields at the second polarizing beam splitter (PBS). The second PBS is a Glan-Thompson type with the extinction ratio better than $10^{-5}$ for completely removing the coupling beams. The probe mode which was initially horizontally polarized, labeled as $H$ in Fig.~\ref{fig1}, was then rotated to vertical polarization with another half-wave plate (HWP) and the two orthogonally polarized probe modes are combined at the second beam displacer (BD). The relative phase between the two modes are set with an AR-coated glass plate, $\phi$ plate, inserted in the setup. The BD based interferometer offers excellent phase stability without active phase locking between the two modes because the two modes are in fact quite close to each other and they share common optical mounts. No appreciable phase change was observed at least for a day. The final half-wave plate (HWP), located after the second BD, is necessary to recover the original polarization state of the optical field at the input of the first BD. Finally, the polarization state of the retrieved probe field was analyzed  with a set of half-wave and quarter-wave plates (WP) and a polarizer.

For demonstrating the scheme experimentally, a 795 nm external cavity diode laser (ECDL) and two acousto-optic modulators (AOM) are used to generate the probe and coupling fields which are 3.035 GHz apart, equal to the hyperfine splitting of the two ground states of \textsuperscript{85}Rb. The probe beam and the coupling beam are obtained by diffracting the ECDL output at the 1.5 GHz AOM in the double-pass configuration and at the 80 MHz AOM, respectively \cite{thermal}. Typical peak powers of the probe and the coupling beams were 20 $\mu$W and 1.2 $\sim$ 1.6 mW, respectively, for each path.

%%%%%%%%%%%%%%%%%%%%%%%%%%%%%%%%%%%%%%%%%%%%%%%%%%%%%%%%%

\section{Experimental Results}

%\subsection{EIT spectra measurement}

%%%%%%%%%%%%%%%%%%
\begin{figure}[tp]
\centering
\includegraphics[width=4.5in]{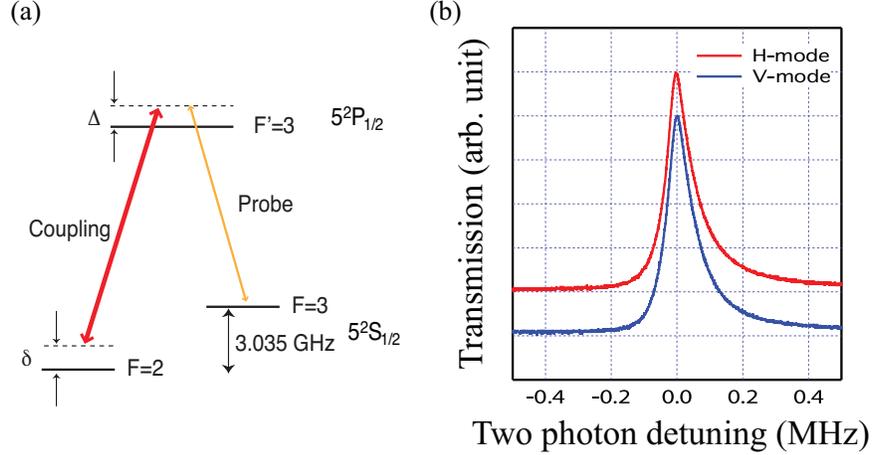}
\caption{(a) The energy level configuration of Rubidium 85. (b) Measurement of the EIT spectra for the two spatial modes, the $H$ labeled and the $V$ labeled modes in Fig.~\ref{fig1}(a), by detuning the coupling frequency. The two EIT spectra are matched at FWHM linewidths of 90 kHz. Note that the spectra are vertically shifted for ease of comparison} \label{fig2}
\end{figure}
%%%%%%%%%%%%%%%%%%%%%%

The atomic vapor quantum memory for the photonic polarization qubit is realized by the means of electromagnetically-induced transparency using the \textsuperscript{85}Rb D1 transition line. The atomic energy level for the EIT configuration is shown in Fig.~\ref{fig2}(a). The probe field and coupling field are resonant with $5^{2}S_{1/2}$ $F=3$ $\rightarrow$ $5^{2}P_{1/2}$ $F'=3$ and $5^{2}S_{1/2}$ $F=2$ $\rightarrow$ $5^{2}P_{1/2}$ $F'=3$ transitions, respectively. The frequency of the probe field is blue-detuned by $\Delta \approx 100$ MHz to obtain the better transmission for the storage and retrieval experiment and the coupling field is set in the two-photon resonance. 

In the Doppler broadened medium as in the warm atomic vapor, the EIT resonance is strongly dependent on many parameters such as the coupling power, the external magnetic field, temperature, and the angle between probe and coupling fields \cite{shuker}. Since the linewidth of EIT resonance is relevant to the decoherence rate, the retrieval efficiency of the light storage and retrieval process decays faster as the EIT linewidth gets broader \cite{figueroa}. For an ideal atomic vapor quantum memory for the photonic polarization qubit, the EIT linewidths for two spatially separated atomic ensembles should be identical. (The polarization state of the retrieved probe field will be biased, if otherwise.) We, therefore, carefully aligned the coupling beams and adjusted the coupling power so that the EIT spectra from the two atomic ensembles are matched. The full width half maximum (FWHM) EIT resonances were measured to be 90 kHz for both paths, see Fig.~\ref{fig2}(b).

%\subsection{Light storage and retrieval}

%%%%%%%%%%%%%%%%%%
\begin{figure}[tp]
\centering
\includegraphics[width=3.5in]{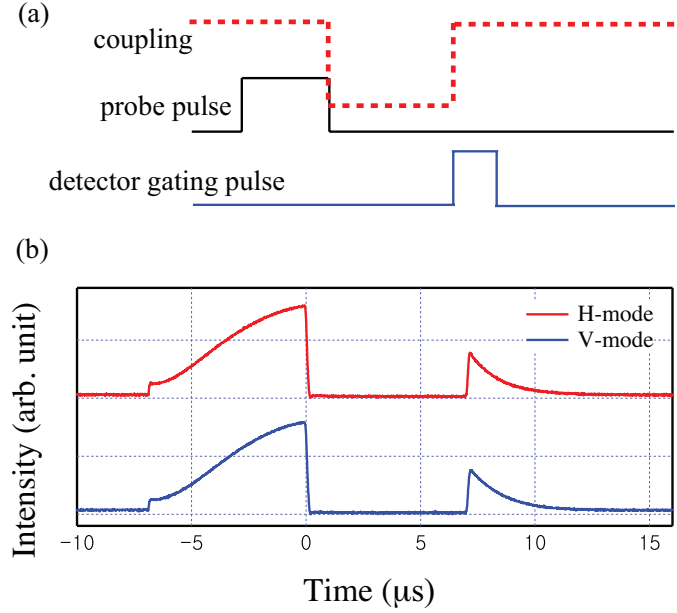}
\caption{(a) Time sequence for the storage and retrieval process. A detector gating pulse is applied to the single-photon detector D for detecting the retrieved probe field only. (b) Storage and retrieval of the $|D\rangle$ polarization state with  7 $\mu$s storage duration. For this measurement, two photocurrents detectors and a PBS are used instead of the state tomography setup.} \label{fig3}
\end{figure}
%%%%%%%%%%%%%%%%%%%%%%

The time sequence for the light storage and retrieval in the EIT medium is shown in Fig~\ref{fig3}(a). We shaped the probe field as a rectangular 7 $\mu$s pulse by turning on and off the 1.5 GHz AOM. The coupling field is initially turned on. After the probe pulse is completely entered into the vapor cell, the coupling beam is turned off. The probe field is, then, stored in atomic collective excitation modes. After some duration of storage time, the photonic polarization qubit is retrieved by turning on the coupling field, see Fig.~\ref{fig3}(b). For the measurement of the retrieved field, a single-photon counting detector was used in the gated mode. (The single-photon detector was single-mode fiber coupled so that additional reduction of probe intensity was not necessary.) The detector is gated for 1 $\mu$s to detect only the retrieved pulse.

%\subsection{Quantum process tomography}

To quantify the performance of the atomic vapor quantum memory for the photonic polarization qubit, we consider the storage-retrieval process as a quantum operation and we analyze the quantum operation on the photonic polarization qubit by performing quantum process tomography \cite{weak}.  Note that a quantum process $\mathcal{E}$ on the quantum state $\rho$ can be represented by a completely-positive linear map as $\mathcal{E}(\rho) =\sum_{mn}  \chi_{mn} E_m \rho E_n^{\dagger}$, where $\chi_{mn}$ represents a positive superoperator, which fully characterizes a quantum process, and the set of $\{E_m\}$ is an operational basis set.

To do the quantum process tomography is to do an experimental reconstruction of the quantum process tomography  matrix $\chi$ and this can be done by analyzing the state of the retrieved polarization qubit for four input states, $|H\rangle$, $|V\rangle$, $|R\rangle= ( |H\rangle-i|V\rangle)/\sqrt{2}$, and $|D\rangle= ( |H\rangle+|V\rangle)/\sqrt{2}$.
The retrieved polarization qubit is analyzed with a set of projection measurements in following basis set:  \{$|H\rangle$, $|V\rangle$, $|R\rangle$, $|L\rangle= ( |H\rangle+i|V\rangle)/\sqrt{2}$, $|D\rangle$, $|A\rangle= ( |H\rangle-|V\rangle)/\sqrt{2}$\}. With these projection measurement results in hand, the quantum process tomography matrix $\chi$, which fully characterizes the quantum process (i.e., the storage and retrieval process) can be determined with the maximum likelihood estimation process \cite{weak}. 

%%%%%%%%%%%%%%%%%%
\begin{figure}[tp]
\centering
\includegraphics[width=4in]{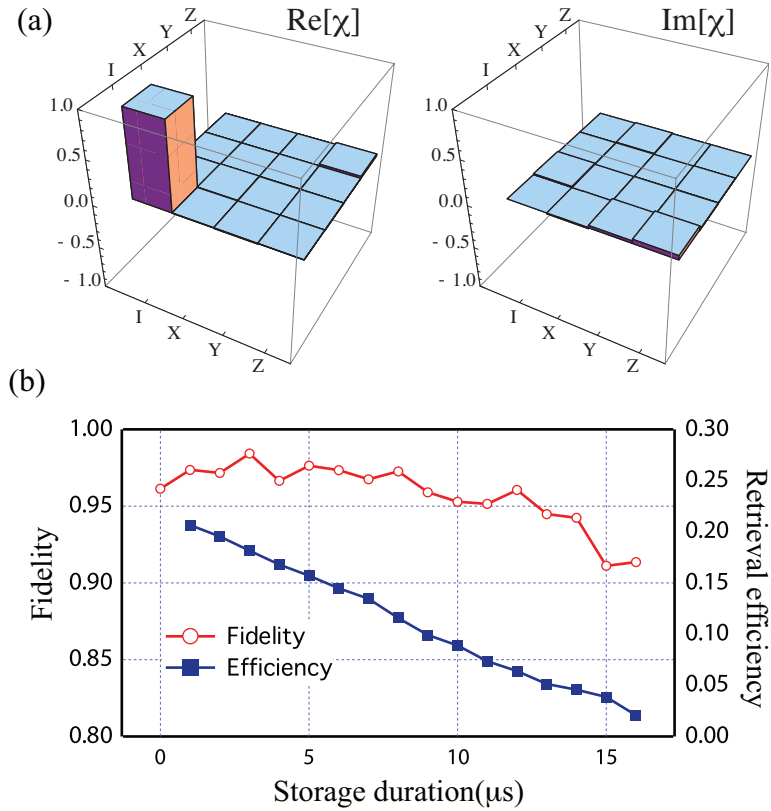}
\caption{(a) Plot of the quantum process tensor $\chi$ for the storage duration of 7 $\mu$s. (b) The fidelity of quantum process and the retrieval efficiency as a function of storage duration. The process fidelity is over 0.91 for up to 16 $\mu$s storage duration.} \label{fig4}
\end{figure}
%%%%%%%%%%%%%%%%%%%%%%

Figure \ref{fig4} shows the result of quantum process tomography. The reconstructed quantum process tomography matrix $\chi$ in the Pauli operator basis $\{I, X, Y, Z\}$ for the storage duration of 7 $\mu$s are represented in Fig.~\ref{fig4}(a). For an ideal quantum memory device, the storage and retrieval process should be represented as an identity operation, i.e the $\chi$ matrix should be peaked only at \{$I,I$\}. As depicted in Fig.~\ref{fig4}(a), the storage and retrieval process operates nearly as an identity operation. 

In order to analyze the quantum process more quantitively, we calculate the quantum process fidelity defined as $F =Tr\left[ \chi_{exp} \chi_{ideal} \right]$, where $\chi_{exp}$ is the experimentally reconstructed quantum process tomography matrix and $\chi_{ideal}$ in this case is the identity operation $I$. In Fig.~\ref{fig4}(b), we show measured process fidelity as well as the retrieval efficiency as a function of the storage duration. The atomic vapor quantum memory for the photonic polarization qubit operates well with high fidelity: the process fidelity over 0.91 for up to 16 $\mu$s storage duration is demonstrated. (We note that the data shown here are not corrected for dark counts and noise counts.) Slight fidelity decrease in this case (from $0$ $\mu$s to $16$ $\mu$s storage duration) is largely caused by the reduced retrieval efficiency which is due to the decoherence of the atomic coherence, thus making contribution from the noise (from the environment and the coupling beam) more significant as the storage duration is increased.

%%%%%%%%%%%%%%%%%%%%%%%%%%%%%%%%%%%%%%%%%%%%%%%%%

\section{Discussion}

Although, in this work, we have tested the polarization qubit quantum memory with the polarization state of a weak laser pulse, our scheme should perform similarly for the single-photon polarization state in principle \cite{kim}. However, for the single-photon probe field, the following problems need to be considered. First, it is necessary to completely remove the coupling field from the probe field. The polarization filtering with a Glan polarizer, used in this work, offers up to $10^{-7}$ extinction ratio and additional filtering can be done, for example, using a multi-pass Fabry-Perot etalon at the extinction ratio of $10^{-5}$ \cite{filter}. Second, there might exists intrinsic background noise photons originated from the spontaneous emission of thermal excited atoms. However, it has been shown that very large signal to noise ratio can in fact be obtained due to the collective enhancement of the atomic ensemble even when the number of stored photon is much less than the number of thermally excited atoms \cite{DLCZ, single1, single2}. Finally, collision-induced fluorescence background noise can limit the performance of the warm vapor qubit memory when the vapor cell is filled with buffer gas for the longer ground-state hyperfine coherence \cite{jiang09, manz07}. We note, however, that the collision-induced fluorescence noise can be greatly suppressed with a buffer-gas-free cell, such as a paraffin-coated vapor cell \cite{jiang09}.

\section{Summary}
We have reported an experimental demonstration of an atomic vapor quantum memory for the photonic polarization qubit.  The atomic quantum memory is realized with the electromagnetically-induced transparency effects of two spatially separated ensembles of warm Rubidium atoms in a single vapor cell. We have also characterized the performance of the atomic vapor quantum memory by performing quantum process tomography which shows that the atomic vapor quantum memory operates as an identity operation with fidelity better than 0.91 for up to 16 $\mu$s storage time. Since the photonic polarization qubit plays an important role in photonic quantum information technologies, we believe that the atomic vapor quantum memory for the photonic polarization qubit (built with a warm atomic vapor cell, rather than cold atoms which require magneto-optical traps) described in this paper will find a wide range of applications in photonic quantum computing and communication research.

%%%%%%%%% Ack %%%%%%%%%%%
\section*{Acknowledgements}
This work was supported by the National Research Foundation of Korea (2009-0070668 and 2009-0084473) and POSTECH BSRI Fund.


\begin{thebibliography}{99}

% Linear optical quantum computing with photonic qubits
% Pieter Kok, W. J. Munro, Kae Nemoto, T. C. Ralph, Jonathan P. Dowling, and G. J. Milburn
\bibitem{kok} P. Kok, W. J. Munro, K. Nemoto, T. C. Ralph, J. P. Dowling, and G. J. Milburn, ``Linear optical quantum computing with photonic qubits," Rev. Mod. Phys. \textbf{79}, 135 (2007)

% A scheme for efficient quantum computation with linear optics
% E. Knill, R. Laflamme, and G. J. Milburn
\bibitem{klm} E. Knill, R. Laflamme, and G. J. Milburn, ``A scheme for efficient quantum computation with linear optics," Nature (London) \textbf{409}, 46 (2001) 

% Long distance quantum communication with atomic ensembles and linear optics
% L.-M. Duan, M. D. Lukin, I. J. Cirac, and P. Zoller
\bibitem{DLCZ} L.-M. Duan, M. D. Lukin, J. I. Cirac, and P. Zoller, ``Long distance quantum communication with atomic ensembles and linear optics," Nature (London) \textbf{414}, 413 (2001)

% Optical quantum memory
% A. I. Lvovsky, B. C. Sanders, and W. Tittel
\bibitem{qm} A. I. Lvovsky, B. C. Sanders, and W. Tittel, ``Optical quantum memory," Nature Photonics  \textbf{3}, 706 (2009)

% Electromagnetically induced transparency: Optics in coherent media
% Michael Fleischhauer, Atac Imamoglu, and Jonathan P. Marangos
\bibitem{eit} M. Fleischhauer, A. Imamoglu, and J. P. Marangos, ``Electromagnetically induced transparency: Optics in coherent media," Rev. Mod. Phys. \textbf{77}, 633 (2005)

% Dark-State Polaritons in Electromagnetically Induced Transparency
% M. Fleischhauer and M. D. Lukin
\bibitem{dark} M. Fleischhauer and M. D. Lukin, ``Dark-State Polaritons in Electromagnetically Induced Transparency," Phys. Rev. Lett. \textbf{84}, 5094 (2000)

% Storage of Light in Atomic Vapor
% D. F. Phillips, A. Fleischhauer, A. Mair, R. L. Walsworth, and M.D. Lukin
\bibitem{Lukin} D. F. Phillips, A. Fleischhauer, A. Mair, R. L. Walsworth, and M. D. Lukin, ``Storage of Light in Atomic Vapor," Phys. Rev. Lett. \textbf{86}, 783 (2001)

% Storage and retrieval of single photons transmitted between remote quantum memories
% T. Chaneliere, D. N. Matsukevich, S. D. Jenkins, S.-Y. Lan, T. A. B. Kennedy, and A. Kuzmich
\bibitem{single1} T. Chaneli\`{e}re, D. N. Matsukevich, S. D. Jenkins, S.-Y. Lan, T. A. B. Kennedy, and A. Kuzmich, ``Storage and retrieval of single photons transmitted between remote quantum memories," Nature (London) \textbf{438}, 833 (2005)

% Electromagnetically induced transparency with tunable single-photon pulses
% M. D. Eisaman, A. Andre, F. Massou, M. Fleischhauer, A. S. Zibrov & M. D. Lukin
\bibitem{single2} M. D. Eisaman, A. Andre, F. Massou, M. Fleischhauer, A. S. Zibrov, M. D. Lukin, ``Electromagnetically induced transparency with tunable single-photon pulses," Nature (London) \textbf{438}, 837 (2005)

% Storage and retrieval of nonclassical photon pairs and conditional single photons generated by the parametric down-conversion process
% K. Akiba, K. Kashiwagi, T. Yonehara, and M. Kozuma
\bibitem{SPDC} K. Akiba, K. Kashiwagi, T. Yonehara, and M. Kozuma, ``Storage and retrieval of nonclassical photon pairs and conditional single photons generated by the parametric down-conversion process," New J. Phys. \textbf{11}, 013049 (2009)

% Quantum memory for squeezed light
% Jurgen Appel, Eden Figueroa, Dmitry Korystov, M. Lobino, and A. I. Lvovsky
\bibitem{squeezed} J. Appel, E. Figueroa, D. Korystov, M. Lobino, and A. I. Lvovsky, ``Quantum memory for squeezed light," Phys. Rev. Lett. \textbf{100}, 093602 (2008)

% Storage and Retrieval of a squeezed vacuum
% Kazuhito Honda, Daisuke Akamatsu, Manabu Arikawa, Yoshihiko Yokoi, Keiichirou Akiba, Satoshi Nagatsuka, Takahito Tanimura, Akira Furusawa, and Mikio Kozuma
\bibitem{honda} K. Honda,  D. Akamatsu, M. Arikawa, Y. Yokoi, K. Akiba, S. Nagatsuka, T. Tanimura, A. Furusawa, and M. Kozuma, ``Storage and Retrieval of a squeezed vacuum," Phys. Rev. Lett. \textbf{100}, 093601 (2008)

% Storage and Retrieval of Thermal Light in Warm Atomic Vapor
%Y.-W. Cho and Y.-H. Kim
\bibitem{thermal} Y.-W. Cho and Y.-H. Kim, ``Storage and retrieval of thermal light in warm atomic vapor," Phys. Rev. A \textbf{82}, 033830 (2010)


% Quantum state transfer between matter and light
% D. matsukevich and A. Kuzmich
\bibitem{mats} D. N. Matsukevich and A. Kuzmich, ``Quantum state transfer between matter and light," Science \textbf{306}, 663 (2004)


% Mapping photonic entanglement into and out of a quantum memory
% K. S. Choi, H Deng, J Laurat, H. J. Kimble
\bibitem{choi} K. S. Choi, H. Deng, J. Laurat, and H. J. Kimble, ``Mapping photonic entanglement into and out of a quantum memory," Nature \textbf{452}, 67 (2008)

% Heralded Single-Magnon Quantum Memory for Photon Polarization States
% H. Tanji, S. Ghosh, J. Simon, B. Bloom, and V. Vuletic
\bibitem{tanji} H. Tanji, S. Ghosh, J. Simon, B. Bloom, and V. Vuleti\'{c}, ``Heralded Single-Magnon Quantum Memory for Photon Polarization States," \prl \textbf{103}, 043601 (2009)


%note
\bibitem{note} The optical depth $d$ is defined as the off-resonant transmittance, $\exp(-d)$, of the EIT spectrum. 
The experiment was done with a relatively `thin' optical medium to avoid unwanted four-wave mixing processes which could occur at higher optical depth, see Ref. \cite{phillips08}. It should be easier to work with a greater optical depth for Rubidium 87 as the ground state hyperfine splitting is much larger than that of Rubidium 85.

% Optimal light storage in atomic vapor
% Nathaniel B. Phillips, Alexey V. Gorshkov, and Irina Novikova
\bibitem{phillips08} N. B. Phillips, A. V. Gorshkov, and I. Novikova, ``Optimal light storage in atomic vapor," Phys. Rev. A \textbf{78}, 023801 (2008)



% Anglular dependence of Dicke-narrowed electromagnetically induced transparency resonances
% M. Shuker, O. Firstenberg, R. Pugatch, A. Ben-Kish, A. Ron, and N. Davison
\bibitem{shuker} M. Shuker, O. Firstenberg, R. Pugatch, A. Ben-Kish, A. Ron, and N. Davison, ``Anglular dependence of Dicke-narrowed electromagnetically induced transparency resonances," \pra \textbf{76}, 023813 (2007)

% Decoherence of electromagnetically induced transparency in atomic vapor
% E. Figueroa, F. Vewinger, J. Appel, and A. I. Lvovsky
\bibitem{figueroa} E. Figueroa, F. Vewinger, J. Appel, and A. I. Lvovsky, ``Decoherence of electromagnetically induced transparency in atomic vapor," Opt. Lett. \textbf{31}, 2625 (2006)




% Reversing the weak quantum measurement for a photonic qubit
\bibitem{weak} Y.-S. Kim, Y.-W. Cho, Y.-S. Ra, and Y.-H. Kim, ``Reversing the weak quantum measurement for a photonic qubit," Opt. Express \textbf{17}, 11978 (2009)

% Quantum teleportation of a polarization state with a complete bell state measurement
\bibitem{kim} Y.-H. Kim, S. P. Kulik, and Y. Shih, `` Quantum teleportation of a polarization state with a complete bell state measurement," \prl \textbf{86}, 1370 (2001)

% Note: An ultranarrow bandpass filter system for single-photon experiments in quantum optics 
\bibitem{filter} D. H\"{o}ckel, E. Martin, and O. Benson, ``Note: An ultranarrow bandpass filter system for single-photon experiments in quantum optics," Rev. Sci. Instrum. \textbf{81}, 026108 (2010)

% Collisional decoherence during writing and reading quantum states
% S. Manz, T. Fernholz, J. Schmiedmayer, and J.-W. Pan
\bibitem{manz07} S. Manz, T. Fernholz, J. Schmiedmayer, and J.-W. Pan, ``Collisional decoherence during wring and reading quantum states," \pra \textbf{75}, 040101(R) (2007)

% Observation of prolonged cohrence time of the collective spin wave of an atomic ensemble in a paraffin-coated \textsuperscript{87}Rb vapor cell
% S. Jiang, X.-M. Luo, L.-Q. Chen, B. Ning, S. Chen, J.-Y. Wang, Z.-P. Zhong, and J.-W. Pan
\bibitem{jiang09} S. Jiang, X.-M. Luo, L.-Q. Chen, B. Ning, S. Chen, J.-Y. Wang, Z.-P. Zhong, and J.-W. Pan, ``Observation of prolonged cohrence time of the collective spin wave of an atomic ensemble in a paraffin-coated \textsuperscript{87}Rb vapor cell," \pra \textbf{80}, 062303 (2009)

\end{thebibliography}
\end{document}